\documentstyle[aps,prb,twocolumn]{revtex}

\def\pdev(#1,#2;#3){\left(
           \kern-0.1em{\partial #1\over\partial #2}
           \kern-0.1em\right)_{\kern-.2em #3}}
\def\batio3{BaTiO$_3$}
\def\ave#1{\langle#1\rangle}
\def\corr(#1,#2){\ave{\Delta{#1}\Delta{#2}}}

\begin{document}

\twocolumn[\hsize\textwidth\columnwidth\hsize\csname
@twocolumnfalse\endcsname

\title{Optimized local modes for lattice dynamical applications}

\author{Jorge \'I\~niguez, Alberto Garc\'{\i}a}
\address{
Departamento de F\'{\i}sica Aplicada II, Universidad del Pa\'{\i}s
Vasco, Apdo. 644, 48080 Bilbao, Spain}

\author{J.M. P\'erez-Mato}
\address{
Departamento de F\'{\i}sica de la Materia Condensada, 
Universidad del Pa\'{\i}s
Vasco, Apdo. 644, 48080 Bilbao, Spain}
\maketitle

\begin{abstract}
We present a new scheme for the construction of highly localized
lattice Wannier functions. The approach is based on a heuristic
criterion for localization and takes the symmetry constraints into
account from the start. We compare the local modes thus obtained with
those generated by other schemes and find that they also provide a
better description of the relevant vibrational subspace.
\end{abstract}

\vskip1pc]

\section{Introduction}

Although the translational symmetry of a crystalline solid imposes a
delocalized basis of Hamiltonian eigenstates (Bloch's functions), it
is sometimes advantageous to consider a transformation to a new set of
basis functions with a local character. Beyond the mathematical
equivalence (both sets span the same space of states), a local
viewpoint is better suited for the analysis of concepts such as
bonding which are eminently local in
character. Recent work on electronic Wannier functions has shown the
usefulness of a local representation in the chemical characterization
of a given band subspace,~\cite{marzari} in the analysis of bonding
topology in a disordered system,~\cite{parrinello} and in more formal
developments.~\cite{other-developments}

The lattice dynamical problem is formally very similar to the
electronic one: a set of Bloch eigenstates (normal modes) represents
the collective vibrations of the atoms in the crystal. A basis change
to a set of local displacement patterns (lattice Wannier functions or
local modes) can in principle be achieved. So far, the main
application for these local modes has been in the field of structural
phase transitions. Typically, the behavior of a given dispersion
branch or set of branches determines the essential instabilities of
the system, and the associated degrees of freedom enter into the
construction of an effective Hamiltonian which reproduces the relevant
physics.  Through the use of a localized basis set, the number of
coupling terms in the effective Hamiltonian can be relatively small,
easing the statistical mechanical treatment and the interpretation of
the results. In particular, the anharmonic terms in the effective
Hamiltonian can be kept local (on-site), in contrast with what
happens in a reciprocal-space description.

This local mode approach has been used extensively in the past to gain
an understanding of the behavior of complex systems, but until
recently the local variables were treated as dummy degrees of freedom
in a semi-empirical model, with their interactions fitted to reproduce
the observed phenomena. In the last few years, a new approach, in
which the effective Hamiltonian is parametrized on the basis of first
principles calculations, has had great success in studies of the phase
transition sequences in perovskite
oxides.~\cite{zhong_batio3,rabe_waghmare,zhong_srtio3} Central to the
parametrization process is the explicit construction of lattice
Wannier functions, and two schemes have been proposed to carry it
out. Zhong, Vanderbilt, and Rabe~\cite{zhong_batio3} (ZVR) used the
structure of the zone-center soft mode in perovskite \batio3 to
construct symmetry-adapted highly localized local modes. Subsequently,
Rabe and Waghmare~\cite{rabe_wannier} (RW) generalized this approach
to reproduce the normal modes at several (typically, high symmetry)
points of the Brillouin Zone.

While both the ZVR and RW approaches have been broadly successful in
the specific problems for which they were conceived, in this paper we
will argue that they are not completely satisfying in some respects.  We
will present a new procedure to generate lattice Wannier functions, an
approach which makes use of the available symmetry information,
produces local modes with a high degree of localization, enables a
systematic improvement of their quality, and is straightforward to
implement.

\section{Method}

We are interested in describing a {\sl relevant subspace} $\cal R$
of the full $3Np$-dimensional configuration space of a crystal with
$p$ atoms per unit cell. Typically, we can choose ${\cal R}$ as a
complete {\sl band} of dispersion branches (complete in the sense that
it is invariant under the action of the space group of the
crystal~\cite{explain-complete}).  Associated to a branch $j$ is a set
of {\sl normal modes} ($3Np$-dimensional vectors) $\{{u}_j^{\bf
k}\}$~\cite{vector-notation} which are eigenvectors of the Fourier
transform of the force-constant matrix.~\cite{force-constant} (The
displacement in the $\alpha$ cartesian direction of the atom $\kappa$
in cell ${\bf l}$ is given explicitly by ${u}_j^{\bf k}({\bf
l},\kappa,\alpha)$.) The normal modes transform according to
irreducible representations of the little groups $G^{\bf k}$. These
representations, considered over the whole BZ, determine the {\sl band
symmetry}. The relevant subspace ${\cal R}$ is spanned by all the
$\{{u}_j^{\bf k}\}$ in the band, but it is clear that any
transformation
\begin{equation}
\tilde u_j^{\bf k} = \sum_{i=1}^n{M_{ji}^{\bf k} {u}_i^{\bf
k}}
\label{eq:bloch_modes}
\end{equation}
will lead to a new basis of extended states which we will
call {\sl Bloch modes}. Here $n$ is the {\sl band dimension}, the
number of dispersion branches in the band.

Having thus specified the relevant subspace by means of Fourier space
variables $\{u_j^{\bf k}\}$, the problem we tackle is the construction of a new
basis $\{w_j^{\bf n}\}$ which is local, as opposed to extended, in
character. Mathematically, the $\bf k$ label should be exchanged by a
local label ${\bf n}$ associated to the different unit cells in the
crystal. Translational symmetry takes the form:
\begin{equation}
{w}_j^{\bf n}({\bf l},\kappa,\alpha) = 
{w}_j^{{\bf n}+{\bf t}}({\bf l}+{\bf t},\kappa,\alpha) \; ,
\label{eq:w_trans}
\end{equation}
which is trivially satisfied by the standard Wannier function
form~\cite{wannier}:
\begin{equation}
{w}_j^{\bf n} = {1\over\Omega}\int_{BZ}
{\exp(-i{\bf kn})\tilde{u}_j^{\bf k}d{\bf k}}
\label{eq:w_def}
\end{equation}
in which $\Omega$ is the volume of the BZ.  A high degree of localization means that the displacement
${w}_j^{\bf n}({\bf l},\kappa,\alpha)$ should be very small or zero
when ${\bf l}$ is a few lattice constants away from ${\bf
n}$. The arbitrariness implicit
in their definition (Eq.~\ref{eq:bloch_modes}) means that the Wannier
functions are non-unique, and a relatively large latitude then exists
to tune their properties. In particular, the degree of localization
has traditionally been the focus of great interest, and recently,
Marzari and Vanderbilt~\cite{marzari} have succeeded in optimizing the
matrices appearing in Eq.~\ref{eq:bloch_modes} to construct
very localized electronic Wannier functions starting from the Bloch
states. A restriction to unitary matrices resulted in an orthonormal
basis of Wannier functions, and the optimization process led to
symmetric-looking functions, even though no symmetry conditions were
explicitly imposed.~\cite{dhv_start} 
In principle, such an approach should work for the vibrational
problem, too. However, we prefer to take an alternate route which 
takes advantage of the knowledge of the band symmetry.

\subsection{Symmetry requirements}

As studied extensively in the literature,~\cite{zak} one should
supplement the translational constraints of Eq.~\ref{eq:w_trans} with
another set of conditions which represent the transformational
properties of the ${w}_j^{\bf n}$ under the effect of the point
symmetry of the crystal. These are most easily discussed by
introducing a symmetry-based definition of the {\sl center} of a mode.
Consider a Wyckoff set with representative site $\bf r$ and the set
$\hat{G}_{\bf r}$ of operations in $G$ that leave $\bf r$
invariant.~\cite{site-symmetry-group}  Given an irreducible
representation $\tau$ of $\hat{G}_{\bf r}$ with dimension $d_{\tau}$,
any $d_{\tau}$ displacement patterns transforming with $\tau$ under
the action of $\hat{G}_{\bf r}$ are said to be centered in $\bf r$.
It is then notationally more convenient to use a double index to label
these patterns: ${w}_{{\bf r}, s}$ where $s$ ranges from 1 to
$d_{\tau}$. The action of the elements of the space group $G$ on this
set generates images at the rest of the positions in the Wyckoff set,
i.e., $d_rd_{\tau}$ patterns $w^{\bf n}_{{\bf r}_i,s}$ per cell, where
$i$ ranges from 1 to $d_r$ (the multiplicity of the Wyckoff set).
This set of lattice functions is represented by the pair $(\bf
r,\,\tau)$ and define a representation of $G$ which is called {\sl
band representation}.~\cite{zak}

A necessary condition for the description of a relevant band subspace
by means of these symmetry-adapted local modes is the equivalence of
the band symmetry of $\cal R$ and the band representation $(\bf
r,\,\tau)$ (in particular this implies $n\,=\,d_rd_{\tau}$).  More
details about the choice of the correct $(\bf r,\,\tau)$ for a given
$\cal R$ are presented in the Appendix, where we also discuss the
transformation properties of the corresponding $\{\tilde u_j^{\bf
k}\}$. Incidentally, since Eq.~\ref{eq:w_def} establishes a
correspondence between lattice Wannier functions and Bloch modes, in
what follows the latter can be also labeled by the site and
representation indexes: $\tilde{u}^{\bf k}_{{\bf r},s}$

\subsection{Practical criterion for localization}

A straightforward scheme to obtain lattice Wannier
functions can be based on a direct use of Eq.~\ref{eq:w_def},
performing the BZ sum by means of any of the standard ``special
k-points'' methods.~\cite{chadi-cohen,monkhorst-pack} The quality of
the subspace description can thus be
systematically improved by simply using denser k-point sets. This
approach can incorporate information about the normal modes throughout
the whole Brillouin zone, as opposed to at just one point (as in the
ZVR method), or at a very special set of high-symmetry k-points (as in
the RW scheme). 

As stated in the Introduction, it is highly desirable that the local
mode basis functions be as localized as possible, in order to permit
the consideration of only a few coupling terms in the effective
Hamiltonian. From the point of view of real applications, a basis of
Wannier functions which are not localized is not efficient, even if it
spans $\cal R$ perfectly. The form of Eq.~\ref{eq:w_def} suggests a
very simple heuristic criterion to achieve a high degree of
localization for the lattice Wannier functions: choose the $M^{\bf k}$
matrices in such a way that the $\tilde{u}^{\bf k}$ Bloch vectors at
different ${\bf k}$ add their contributions coherently at the center of the
Wannier function. Interference effects can then be counted on to
automatically dampen the amplitude of the displacements at sites away
from the center.

Both the symmetry requirements and the localization condition can be
formulated in the following way.  Assume a $({\bf r}, \tau)$ pair has
been determined on the basis of band symmetry, and that we focus on
the construction of local modes at cell $\bf n$=$\bf 0$. Consider a
set of $d_{\tau}$ ($3Np$-dimensional) orthonormal vectors $\{ x_{{\bf
r},s}\}$ which are centered in $\bf r$, transform with irrep $\tau$
and involve atoms in an orbit as close to $\bf r$ as
possible.~\cite{x-orbit} The localization criterion is implemented by
requiring that $\tilde{u}^{\bf k}_{{\bf r},s}$ be orthogonal to
$x_{{\bf r},t}$ if $s\neq t$, and ``parallel'' (meaning that their
scalar product is positive) if $s=t$.~\cite{x-criterion}  It can be
seen that this condition fixes the form of the $M^{\bf k}$ matrices,
up to overall normalization factors.  In general, the $M^{\bf k}$ will
not be unitary, with the result that two lattice Wannier functions at
different cells $w_{{\bf r},s}^{\bf n}$ and ${w}_{{\bf r}',s'}^{\bf
n'}$ will not be orthogonal if the pairs $({\bf r},s)$ and $({\bf
r}',s')$ are not equal.~\cite{non-orthog} In the next section we will
provide a simple worked example of the new construction scheme and
will compare its results to those of other methods.

\section{Examples and Discussion}

In order to illustrate the scheme presented in the previous section,
we will employ a two-dimensional model crystal with two different
atoms which occupy the $1a$ (white) and $1b$ (black) Wyckoff positions
of the plane group $p4mm$ (See Fig.~\ref{fig:crystal} a)). A simple
harmonic model for the force constants (with the couplings among the
white atoms considered up to fourth nearest neighbors and the rest to
first nearest neighbors, which corresponds to 6 independent  
parameters) gives the dispersion branches of panel b) in
the figure. We will focus our attention on the two optical branches,
which form a single band since they are essentially degenerate at the
$\Gamma$ and M points. These optical branches transform according to
the decompositions
\begin{equation}
\begin{array}{lll}
\Gamma&(4mm):& E\cr
{\rm X}&(2mm):&B_1+B_2\cr
{\rm M}&(4mm):&E \;,
\end{array}
\label{eq:bz_map}
\end{equation}
in irreducible representations of the little co-groups at the high
symmetry points. A simple application of the procedure spelled out in
the Appendix shows that the band representation compatible with
the above band symmetry is that represented by the pair $(\circ,E)$,
in which $E$ is a two-dimensional irreducible representation which
turns out to be the vector representation of the point symmetry group
at $\circ$. The set of $\{x\}$ vectors is then trivial
to construct: as the ``$\circ$'' Wyckoff position is occupied, it is
just enough to make $x_{{\circ},1}$ and $x_{{\circ},2}$ unit
vectors attached to the central atom and pointing in the $x$ and
$y$ cartesian directions, respectively. For this crystal structure, the
simplest non-trivial set of special k-points is given by
$\{(1/8,1/8);(1/8,3/8);(3/8,3/8)\}$. The explicit application of the
localization criterion proceeds as follows. At each k-point in the set
the normal modes are computed and the $M^{\bf k}$ matrices constructed. For
example, at $(1/8,3/8)$, the normal modes are
\begin{equation}
\begin{array}{ll}
{u}^{\bf k}_{1} = &(\ldots;0.23,0.93;\ldots)\cr
{u}^{\bf k}_{2} = &(\ldots;0.84,-0.19;\ldots)\; ,
\end{array}
\label{eq:ex_modes}
\end{equation}
where the ``$\ldots$'' refer to displacements on atoms other than the
one at the center. The ``coherent addition at the center'' condition
then becomes:
\begin{equation}
\begin{array}{l}
0.23\,M_{11}+0.84\,M_{12} > 0\cr
0.23\,M_{21}+0.84\,M_{22} = 0\cr
0.93\,M_{11}-0.19\,M_{12} = 0\cr
0.93\,M_{21}-0.19\,M_{22} > 0\; ,
\end{array}
\label{eq:m_cond}
\end{equation}
and is satisfied by
\begin{equation}
M=\left(\matrix{0.200&0.980\cr
                0.964&-0.264\cr}\right) \; ,
\label{eq:m_matrix}
\end{equation}
uniquely defined but for row-specific arbitrary factors. Since M is
not unitary, the two optical Bloch vectors $\tilde u_{\circ,s}$ at
this k-point will not be orthogonal (although they can of course still
be chosen to be normalized).

Once this procedure has been performed at every k-point in the set,
the integral (sum) in Eq.~\ref{eq:w_def} can be carried out to give
the components of the lattice Wannier functions. Since the $\{\tilde
u^{\bf k}\}$ determined by the localization criterion also satisfy the
symmetry compatibility relations (see Appendix), the Wannier functions
are symmetry-adapted. In Fig.~\ref{fig:shells}~c) we show the
displacements associated to the local mode $w_{\circ,1}$, which
transform as the first component of the vector representation $(E)$ of
$4mm$. The degree of localization of these lattice Wannier functions
can be gauged by computing the contribution to the total norm from a
given shell around the center atom, as presented on
Table~\ref{tab:shells}. Less than one per cent of the norm is outside
the fourth shell (which corresponds roughly to the second-neighbor unit
cells).  If the integral in Eq.~\ref{eq:w_def} is computed using a
denser special-point set, the degree of localization is maintained, as
can be seen by comparing the columns labeled ``this work (3 k)'' and
``this work (10 k)'' on Table~\ref{tab:shells}. This means that the
quality of the local modes can be systematically improved while
retaining a high degree of localization.

It is enlightening to compare this scheme to that of Rabe and
Waghmare.~\cite{rabe_wannier} In the latter the analysis of the
symmetry compatibility relations proceeds in the same way, and once
the right $({\bf r},\tau)$ set has been identified, a series of
orthonormal ${x}$ sets is constructed at successive shells
centered on $\bf r$. The extent of the outermost shell fixes the
localization of the Wannier functions by construction, and the actual
atomic displacements are determined by fitting to the normal modes
computed at a few high-symmetry points of the Brillouin Zone. In
essence, the normal-mode information determines the weight assigned to
each symmetry-adapted shell, so there is a tradeoff between the extent
of the lattice Wannier functions and the amount of information from
the real dispersion relations that can be used in the construction
procedure. For example, in the PbTiO$_3$ work, Rabe and Waghmare found
that adding information about the normal modes at the X point
resulted in a less localized local mode than if only the
$(\Gamma,{\rm M},{\rm R})$ set was used.~\cite{rabe_waghmare}  In contrast, our
scheme can deal with the extra k-point without loss of localization: our
local modes for PbTiO$_3$ using four high-symmetry
points~\cite{mp-unshifted} are more localized than the best (three
point) RW lattice Wannier functions.

It is clear that localization cannot be the main quality
criterion for the construction of local modes. If it were, then the
ZVR scheme, which uses only one high-symmetry k-point
to construct a (very localized) lattice Wannier
function, would be the method of choice. In fact, the real test for
local mode sets is the degree to which they reproduce the energetics
of the relevant subspace $\cal R$. 
That is, in our case, the degree to which the
dispersion relations of the effective Hamiltonian 
\begin{equation}
H_{\rm eff}=H_{\rm eff}(Q_1,Q_2,\ldots,Q_{Nn})
\label{eq:heff}
\end{equation}
match the real dispersion branches associated with $\cal R$.

In Eq.~\ref{eq:heff}, the variables $Q_i$ are the amplitudes of the
local mode variables, so that $H_{\rm eff}$ can be
thought of as the ``projection'' of the complete Hamiltonian into
the relevant subspace $\cal R$ 
(which is typically considered as energetically decoupled
from the rest of the configuration space of the crystal).  
The explicit form of $H_{\rm eff}$ will
depend on the detailed structure of the lattice Wannier functions. In
particular, the number of distinct coupling coefficients (representing
the interaction of modes at different sites) which one
should take into account in $H_{\rm eff}$ is determined by the spatial
extent of the local modes.

We have constructed effective Hamiltonians for the model crystal for
each of the three local-mode construction schemes discussed above (we
obtain the coupling between $w^{\bf n}_{\circ,s}$ and $w^{{\bf
n}'}_{\circ,s'}$ by calculating the energy associated to the crystal
when it is distorted by just these modes). The original crystal Hamiltonian
involved interactions up to fourth nearest neighbors for white
atoms. Since the local modes involve basically displacements of the
central white atom, we have kept the same range of interaction in
$H_{\rm eff}$, but now referring of course to fourth nearest local
modes. This amounts to using ten independent coupling coefficients; a
larger number of parameters would not be reasonable in a practical
application.

ZVR-style local modes are very localized and do not couple beyond the
fourth neighbor shell, so the considered $H_{\rm eff}$ includes all
the existing interactions.  This can be seen on panel a) of
Figure~\ref{fig:gm}: the dispersion branches computed from $H_{\rm
eff}$ match the exact ones at the $\Gamma$ point. However, $H_{\rm
eff}$ gives a poor description of the dispersion branches away from
$\Gamma$, as it should be expected in view of the construction
procedure. (Incidentally, the inverse of Eq.~\ref{eq:w_def} leads to
Bloch modes which are not normalized to unity, except at the $\Gamma$
point. The standard analysis of $H_{\rm eff}$ as given would lead to the
low-lying dispersion branches in the figure. The higher
branches are obtained by considering the corresponding generalized
eigenvalue problem.) Panel b) shows that the $H_{\rm eff}$ constructed
on the basis of RW local modes gives a good qualitative overall
description of the dispersion, but fails to match the exact branches
at the $\Gamma$ point (as it should, given that this point was used in
the construction scheme). The reason is that the local modes are more
extended, and it is necessary to include couplings to further shells
(at least up to seventh nearest neighbors)~\cite{more-info} for the
match to be essentially perfect. This means that the RW scheme does
not lead to efficient local modes, in the sense stated
above.~\cite{rabe-fit} The situation gets worse if more accuracy is
needed in the overall description of the dispersion branches: the
local modes turn out to be more extended, and even more coupling terms
are needed in $H_{\rm eff}$.

In contrast, the local modes constructed following our heuristic
criterion for localization do exhibit good efficiency (the dispersion
branches do not change much when couplings to more than fourth nearest
neighbors are included) and provide a very good qualitative match of
the true branches throughout the BZ (Fig.~\ref{fig:gm} c)). (It should
be noted that our construction scheme does not involve any
high-symmetry points, hence the offset of the branches at
$\Gamma$ and M. We trade an overall good match for perfect accuracy
at a few points.)  Since a few coupling terms are enough to take into
account the structure of the local modes, and the resulting $H_{\rm
eff}$ provides a good fit to the true branches, our lattice Wannier
functions are well suited for the local representation of the relevant
subspace $\cal R$. Moreover, they can be improved if needed by
including more k-points in the integration set, with only a minor
sacrifice in the compactness of the effective Hamiltonian.

We find these general conclusions to remain valid when more complicated
interaction models are considered.

The practical application of our method of local mode construction to
real materials requires the knowledge of the normal modes at general
points of the Brillouin Zone. This information is easily obtained with
modern linear-response codes without the need for large supercells.
In the field of phase transitions, the use of this new scheme should
enable the study of more complicated situations than those considered
up to now. Competition of instabilities associated to different
regions of the BZ or complications derived from anti-crossing
phenomena are examples in which this method is bound to be useful.  On
the other hand, this work might provide an illustration of some of its
theoretical underpinnings: the physical interpretation of the band
representation associated to a dispersion
band~\cite{band-interpretation} or the symmetry-induced continuity of
phonon spectra~\cite{michel-walker-zak} are two instances of this.

\section{Conclusions}

We have presented a straightforward scheme for the construction of
very localized lattice Wannier functions, with explicit consideration
of crystal symmetry.  The new localization procedure enables a
systematic improvement in the description of the relevant physics (by
simply using denser sets of special k-points in the BZ integration)
while still being quite efficient in regard to the number of coupling
parameters needed in the effective Hamiltonian. Besides, the present
method is straightforward to implement.

\section*{Acknowledgements}

We thank Karin Rabe, Philippe Ghosez, and David Vanderbilt for useful
comments.  This work was supported in part by the UPV research grant
060.310-EA149/95 and by the Spanish Ministry of Education grant
PB97-0598. J.I. acknowledges fellowship support from the Basque
regional government and thanks Agustin V\'algoma for comments on the
manuscript.

\section*{Appendix}

Let us study in more detail the equivalence between the {\sl band symmetry} 
emerging from the transformation properties of the Bloch modes and the 
{\sl band representation} associated to a $({\bf r},\tau)$ set.~\cite{zak}
This can be done by considering the action of $G$ on the 
$({\bf r},\tau)$ set of local modes and, consequently, on the
associated Bloch modes $\tilde{u}^{\bf k}_{{\bf r}_i,s}$. 

In order to proceed, we need to formulate the transformation
properties of the modes ${w}^{\bf 0}_{{\bf r},s}$ under the action of
$\{\bar{R}|\bar{\bf v}\}\in \hat{G}_r$.  Since we consider
$\{\bar{R}|\bar{\bf v}\}$ acting on the modes themselves
and not on their components, we denote symmetry operations by the associated
operators $O\{\bar{R}|\bar{\bf v}\}$.  We have
\begin{equation}
O\{\bar{R}|\bar{\bf v}\}\,{w}^{\bf 0}_{{\bf r},s}\;
=\;
\sum_{h=1}^{d_{\tau}}\,D^{\tau}_{hs}(\{\bar{R}|\bar{\bf v}\})
\,{w}^{\bf 0}_{{\bf r},h}
\label{local-transformation}
\end{equation}
where ${\bf D}^{\tau}(\{\bar{R}|\bar{\bf v}\})$ is the matrix associated to 
$\{\bar{R}|\bar{\bf v}\}$ by irrep $\tau$. 
Now, we consider the rest of elements in the $({\bf r},\tau)$ set. 
They can be mathematically defined as
\begin{equation}
{w}^{\bf n}_{{\bf r}_i,s}\;
:=\;
O\{E|{\bf n}\}\,O\{R_i|{\bf v}_i\}\,{w}^{\bf 0}_{{\bf r},s}
\end{equation}
where $\{E|{\bf n}\}$ is a lattice translation and $\{R_i|{\bf v}_i\}$ is 
one of the $d_r$ elements in $G/G_r$, which are chosen so that all the 
${\bf r}_i:=\{R_i|{\bf v}_i\}{\bf r}$ lie in the same cell. 
The action of any $\{R|{\bf v}\}\in G$ on an arbitrary 
${w}^{\bf n}_{{\bf r}_i,s}$ can be decomposed in: a lattice translation, 
a change of the center  
and a local transformation. Mathematically, this is expressed as
\begin{eqnarray}
O\{R|{\bf v}\}\,(O\{E|{\bf n}\}\,O\{R_i|{\bf v}_i\}\,
{w}^{\bf 0}_{{\bf r},s})\;
=\cr
O\{E|\{R|{\bf v}\}({\bf n}+{\bf r}_i)-{\bf r}_j\}\,
O\{R_j|{\bf v}_j\}\,
O\{\bar{R}|\bar{\bf v}\}\,{w}^{\bf 0}_{{\bf r},s}
\end{eqnarray}
where $\{R_j|{\bf v}_j\}\in G/G_r$ and $\{\bar{R}|\bar{\bf v}\}\in
\hat{G}_{\bf r}$ are univocally determined. 
Together with Eq.~\ref{local-transformation}, this expression defines the 
band representation and, by using the inverse of Eq.~\ref{eq:w_def}, 
it can be written in the basis of Bloch modes. 
We obtain
\begin{eqnarray}
O\{R|{\bf v}\}\,\tilde{u}^{\bf k}_{{\bf r}_i,s}\;
=\cr
\exp{(-i\,R{\bf k}\,(\{R|{\bf v}\}{\bf r}_i-{\bf r}_j))}\;
\sum_{h=1}^{d_{\tau}}\,D^{\tau}_{hs}(\{\bar{R}|\bar{\bf v}\})\,
\tilde{u}^{R{\bf k}}_{{\bf r}_j,h}
\label{band-representation}
\end{eqnarray}
By examining the representations this equation defines in the high
symmetry k-stars, it can be easily checked whether the band representation
$({\bf r},\tau)$ is equivalent to the band symmetry we want to
describe.~\cite{bacry-michel-zak}

Once a convenient $({\bf r},\tau)$ set is chosen,
Eq.~\ref{band-representation} fixes the requirements on Bloch modes so
that they lead to symmetry adapted local modes $w^{\bf n}_{{\bf
r}_i,s}$.  For pure translations, Eq.~\ref{band-representation}
reduces to Bloch theorem.  Point symmetry determines the
transformation properties of the $d_{\tau}d_r$ Bloch modes in each
k-point and establishes the relationship of these with those in the
rest of the k-star.  However, Eq.~\ref{band-representation} does not
determine the form of the $M^{\bf k}$ matrices completely.  For instance, in a
general k-star ($\bar{G}^{\bf k}=\{E\}$) no condition is imposed on
the choice of Bloch modes in a representative ${\bf k}$, though, once
this is done, the modes in the rest of the star are fixed.  This is
the freedom we use in our construction procedure to get the
localization of the modes.

\begin{table}
\caption{Contribution to the local mode's norm from each of the first
four shells around the $\circ$ atom. For RW, the complete mode is
contained in the first four shells, but it is not normalized to
unity. For local modes calculated with our scheme using 3 and 10
k-points, less than one per cent of the norm is outside these four
shells.}
\label{tab:shells}
\begin{tabular}{lccc}
Shell& This work (3 k)& This work (10 k)&RW\\
\tableline
1st&0.8828&0.8817&0.8241\\
2nd&0.1100&0.1099&0.1617\\
3rd&0.0001&0.0006&0.0029\\
4th&0.0009&0.0014&0.0021\\
\tableline
Total (1st to 4th)&0.9938&0.9936&0.9908\\
Complete mode&1.0000&1.0000&0.9908\\
\tableline
\end{tabular}
\end{table}

\begin{figure}
\caption{a) Two-dimensional model crystal. b) Dispersion branches
for the model crystal assuming a simple harmonic model.}
\label{fig:crystal}
\end{figure}

\begin{figure}
\caption{Lattice Wannier functions for the optical band of the model
crystal. Only the function transforming as the first component of
the two-dimensional representation $E$ of $4mm$ is shown on each
panel. a) ZVR method using information at $\Gamma$;  b) Rabe-Waghmare
method using information at $(\Gamma,{\rm X},{\rm M})$; c) This work, using a
three k-point special set. For c), displacements can be calculated for
up to six shells around the $\circ$ atom.}

\label{fig:shells}
\end{figure}

\begin{figure}
\caption{Optical dispersion branches for the model crystal in the
$\Gamma$-M direction. Exact results (lines) are compared to those
from effective Hamiltonians (squares) corresponding to: a) ZVR local
modes; b) RW lattice Wannier functions; c) This work's local modes
(obtained with a 3 k-point set).  In a) we also present (as circles)
the branches obtained if the generalized eigenvalue problem is not
considered (see text).}
\label{fig:gm}
\end{figure}


\begin{references}
\bibitem{other-developments} In particular, there is an interesting
connection to the modern theory of polarization, and the local
viewpoint is particularly relevant in ``order-N'' schemes. (See
Ref.~\protect\onlinecite{marzari} and references therein.)
\bibitem{explain-complete} In practice, rather than using actual
dispersion branches of the crystal, $\cal R$ could be constructed 
using a physically motivated criterion, such as the avoidance of
irrelevant anti-crossings.
\bibitem{vector-notation} We reserve the bold typeface for
three-dimensional vectors.
\bibitem{force-constant} We do not solve the equations of motion of
the system so, strictly speaking, the $u^{\bf k}$ are not dynamical normal
modes.
\bibitem{dhv_start} But that result might not be always guaranteed.
Note that, in some cases, the functions used as a starting
point for the minimization were chosen to display the appropriate
symmetry.
\bibitem{non-orthog} This loss of orthogonality, also present in the
ZVR and RW schemes, is not really serious for practical applications.
\bibitem{mp-unshifted} The four high-symmetry points can be seen as a
particular case of a special-point set (an unshifted 222
Monkhorst-Pack grid).
\bibitem{more-info} Complete details and graphs about the convergence
with the number of neighbors can be obtained from the authors upon
request.
\bibitem{rabe-fit} The authors of Ref.~\onlinecite{rabe_waghmare} used
a different procedure to construct the $H_{\rm eff}$, imposing the
condition that it exactly reproduces the eigenvalues at selected
(high-symmetry) k-points. We feel that our approach is more appropriate
to check the quality of the local modes, since the effective
Hamiltonians do not depend on the choice of a concrete set of k
points.
\end{references}
\end{document}